\documentclass[12pt]{article}
\usepackage{amsmath,amssymb,amsfonts,epsfig}
\usepackage{graphics}
\usepackage{amssymb,amsmath}
\usepackage{cite}
\usepackage[usenames]{color}
\usepackage[usenames,dvipsnames]{pstricks}
\usepackage{epsfig}
\usepackage{pst-grad} 
\usepackage{pst-plot} 
\newcommand{\ct}{\cite}
\newcommand{\lb}{\label}

\newcommand{\bc}{\begin{center}}
\newcommand{\ec}{\end{center}}
\newcommand{\bd}{\begin{displaymath}}
\newcommand{\ed}{\end{displaymath}}
\newcommand{\be}{\begin{equation}}
\newcommand{\ee}{\end{equation}}
\newcommand{\ba}{\begin{array}}
\newcommand{\ea}{\end{array}}
\newcommand{\bea}{\begin{eqnarray}}
\newcommand{\eea}{\end{eqnarray}}
\newcommand{\bt}{\begin{tabular}}
\newcommand{\et}{\end{tabular}}

\newcommand{\bp}{\begin{picture}}
\newcommand{\ep}{\end{picture}}
\newcommand{\bfi}{\begin{figure}}
\newcommand{\efi}{\end{figure}}

\begin{document}

\begin{titlepage}

\begin{center}
{{\Large {\bf Mirror or Superstring-Inspired Hidden Sector of the
Universe, Dark Matter and Dark Energy} }}
\vspace*{10mm}


{\bf \large{ C. R. Das $^{a}$,  L. V. Laperashvili $^{b}$,
H.B.~Nielsen $^{c}$ and A. Tureanu $^{d}$}}
\end{center}
\begin{center}
\vspace*{0.4cm} {\it { $ ^a$\large Centre for Theoretical Particle
Physics, Technical University of Lisbon, \\ Avenida Rovisco Pais, 1
1049-001 Lisbon,
Portugal\\
$^b$ The Institute of Theoretical and Experimental Physics,\\
Bolshaya Cheremushkinskaya, 25, 117218 Moscow, Russia \\
$^c$ Niels Bohr Institute, Blegdamsvej 17-21, DK-2100 Copenhagen, Denmark \\
$^d$ Department of Physics, University of Helsinki\\ and Helsinki
Institute of Physics, P.O.Box 64, FIN-00014 Helsinki, Finland}}\\
\vskip 0.2cm \centerline{\tt crdas@cftp.ist.utl.pt, laper@itep.ru,
hbech@nbi.dk, anca.tureanu@helsinki.fi}

\vspace*{1.0cm}
\end{center}

\end{titlepage}

\begin{center}{\bf Abstract}\end{center}
\begin{quote}

We develop a concept of parallel existence of the ordinary (O) and
hidden (H) worlds. We compare the two cases: 1) when the hidden
sector of the Universe is a mirror counterpart of the ordinary
world, and 2) when it is a superstring-inspired shadow world
described, in contrast to the mirror world, by a symmetry group
(or by a chain of groups), which does not coincide with the
ordinary world symmetry group. We construct a cosmological model
assuming the existence of the superstring-inspired $E_6$
unification, broken at the early stage of the Universe into
$SO(10)\times U(1)_Z$ -- into the O-world, and $SU(6)'\times
SU(2)'_{\theta}$ -- into the H-world. As a result, we obtain the
low energy symmetry group $G'_{SM}\times SU(2)'_{\theta}$ in the
shadow world, instead of the Standard Model group $G_{SM}$
existing in the O-world. The additional non-Abelian
$SU(2)'_{\theta}$ group with massless gauge fields, ''thetons", is
responsible for dark energy. Considering a quintessence model of
cosmology with an inflaton $\sigma$ and an axion $a_{\theta}$,
which is a pseudo Nambu-Goldstone boson induced by
$SU(2)'_{\theta}$-group anomaly, we explain the origin of dark
energy, dark matter and ordinary matter. In the present model we
review all cosmological epochs (inflation, reheating,
recombination and nucleosynthesis), and give our version of the
baryogenesis. The cosmological constant problem is also briefly
discussed.

\end{quote}
\vspace*{1.0cm}

\section{Introduction}

In this paper we have presented the hypothesis that there may
exist in the Universe the ordinary (O) and hidden (H) worlds
assuming the existence of the mirror (M) or superstring-inspired
shadow (H) counterpart of our observable O-world. Constructing a
new cosmological model with superstring-inspired $E_6$ unification
in the 4-dimensional space, which is broken at the early stage of
the Universe into $SO(10)\times U(1)_Z$ -- in the O-world, and
$SU(6)'\times SU(2)'_{\theta}$ -- in the H-world, we try to
explain the origin of the Dark Energy (DE), Dark Matter (DM) and
Ordinary Matter (OM), in accordance with energy densities given by
recent cosmological observations. The model describes the
inflation, reheating, baryogenesis and nucleosynthesis epochs of
our Universe, confirming the $\Lambda CDM$ model of our
accelerating Universe with a tiny value of the cosmological
constant (CC), $\Lambda$.

The study is based on Refs.~\ct{1,2} and presents a development of
the ideas considered previously in Refs.~\ct{3}. However, in
present investigation we give a new interpretation of the possible
accelerating expansion of the Universe, as far as inflation and
baryogenesis.

\subsection{ Recent results of cosmological and astrophysical observations}

Modern models for DE and DM are based on precise measurements in
cosmological and astrophysical observations \ct{8,9,14}.

For the present epoch, the Hubble parameter $H=H_0$ is given by
the following value \ct{8}:
\be H_0 = 1.5 \times 10^{-42}\,\,{\rm{GeV}}, \lb{1AC} \ee
and the critical density of the Universe is
\be \rho_{c} = 3H^2/8\pi G = {(2.5\times 10^{-12}\,\,
{\rm{GeV}})}^4, \lb{2AC} \ee
where $G$ is the gravitational constant.

Cosmological measurements give the following density ratios of the
total Universe \ct{8}:
\be \Omega = \Omega_r + \Omega_m+ \Omega_\Lambda \approx 1,
\lb{3AC} \ee
where $\Omega_r \ll 1$ is a relativistic (radiation) density
ratio, and
\be \Omega_{\Lambda} = \Omega_{DE}\approx 75\%\, \lb{4AC} \ee
for the mysterious DE, which is responsible for the acceleration
of the Universe. The total matter density is
\be
 \Omega_m\approx \Omega_M + \Omega_{DM} \approx 25\%, \lb{5AC} \ee
with
\be \Omega_M \approx \Omega_B \approx 4\%       \lb{6AC} \ee
- for (visible) ordinary matter and baryons, while
\be \Omega_{DM} \approx 21\%          \lb{7AC} \ee
- for the Dark Matter (DM). These results give:
\be \Omega_{DM}/\Omega_B \approx 5 \lb{8AC}. \ee
The $\Lambda CDM$-cosmological model \ct{17} predicts that the
cosmological constant CC is
\be \Lambda = 8\pi G \rho_{vac}^{(eff)},  \lb{9AC} \ee
where the value $\rho_{vac}^{(eff)}$ is the effective vacuum
energy density of the Universe, which coincides with $\rho_{DE}$.
Using Eqs.~(\ref{2AC}) and (\ref{4AC}), we can calculate the dark
energy density:
\be \rho_{DE} = \rho_{vac}^{(eff)}\simeq 0.75\rho_c\simeq
(2.3\times 10^{-3}\,\,{\mbox{eV}})^4. \lb{10AC} \ee
This is a result of recent cosmological observations \ct{14},
which also fit the equation of state for DE: $w = p/\rho$ with the
following constant value of $w$:
\be w = -1.05\pm 0.13\,\,{\rm{(statistical)}}\pm 0.09\,\,
{\rm{(systematic)}}. \lb{11AC} \ee
In the units $\kappa=1$, where $\kappa^2=8\pi G$, we have the
cosmological constant:
\be \Lambda = \rho_{DE}\simeq (2.3\times
10^{-3}\,\,{\mbox{eV}})^4, \lb{12AC} \ee
which is extremely small.

This result is consistent with the present model of accelerating
Universe \ct{17} (see also reviews \ct{18}), dominated by a tiny
cosmological constant $\Lambda$, $w=-1$ and Cold Dark Matter (CDM)
-- this is the so-called $\Lambda CDM$ scenario.

\subsection{The main assumptions of the present model}

Our model is based on the following assumptions:

$\bullet$ Grand Unified Theory (GUT) is inspired by the
superstring theory \ct{21,23,26}, which predicts $E_6$ unification
in the 4-dimensional space \ct{26}, occurring at the high energy
scale $M_{E_6}\approx 10^{18}$ GeV.

$\bullet$ There exists a Mirror World (M) \ct{27,28}, which is a
duplication of our Ordinary World (O), or shadow Hidden World (H)
(see Refs.~\ct{29}). H-world is not identical with the O-world
having different symmetry groups.

$\bullet$ DE and DM are described by the mirror world (M) with a
broken mirror parity (MP) (see Refs.~\ct{32,33,36,38,39,41,42}),
or by the superstring-inspired shadow H-world considered in
Refs.~\ct{1,2,3}.

$\bullet$ We assume that $E_6$ unification restores mirror parity
at high energies $\approx 10^{18}$ GeV (and at the early stage of
the Universe). Then the mirror world exists at the scale
$M'_{E6}=M_{E6}\approx 10^{18}$ GeV, and the symmetry group of the
Universe is $E_6\times E'_6$  {\footnote {The superscript 'prime'
denotes the M- or H-world.}.

The paper is organized as follows: In the next section we
introduce the $E_6$ unification in the 4-dimensional space-time
inspired by superstring theory and the breakdown of this
unification by different ways. In Sec. III, we discuss the
hypothesis of the existence in Nature a mirror (M) world parallel
to the visible ordinary (O) world, their particle content, mirror
world with broken mirror parity and seesaw scale in the ordinary
and mirror worlds. In Sec.IV we present the existence of
low-energy symmetry groups $G_{SM}= SU(3)_C\times SU(2)_L\times
U(1)_Y$ in the O-world, and $ G' = SU(3)'_C\times SU(2)'_L\times
SU(2)'_{\theta}\times U(1)'_Y $ in the H-world. The group $G'$ has
an additional non-Abelian group $SU(2)'_{\theta}$ with gauge
fields 'thetons', which are neutral massless vector particles.
These 'thetons' have a macroscopic confinement radius
$1/\Lambda'_{\theta}$, where $\Lambda'_{\theta} \sim 10^{-3}$ eV.
The breaking mechanism of the $E_6$ unification is presented in
Sec.V. It was shown that this breaking is realized with the Higgs
fields $H_{27}$ belonging to the 27-plet of $E_6$ - in the
O-world, and with  $H_{351}$ belonging to the 351-plet of $E'_6$ -
in the H-world. We discuss a problem of walls avoiding an
unacceptable wall dominance. Sections VI-VIII are devoted to the
problem of cosmological constant. We show that the cancellation
between the 'bare' cosmological constant, $\lambda$, and the
vacuum energy stress, $8\pi G\rho_{vac}$, described only by the SM
contributions of the ordinary and hidden worlds, explains the
small value of dark energy density $\rho_{DE} = \Lambda \simeq
(2.3\times 10^{-3}\,\,{\mbox{eV}})^4$ by the condensation of
$\theta$-fields. Inflationary, reheating and radiation epochs of
our Universe are reviewed in Sections IX and X. Inflationary
potential is described by Coleman-Weinberg potential. The ordinary
and hidden sectors of the Universe have different cosmological
evolutions and never have to be in equilibrium with each other.
The Big Bang Nucleosynthesis (BBN), which is considered in Sec.XI,
gives the constraint: $T'<T$, where $T(T')$ is O-(H-) temperature
of the Universe. The difference between the O- and H-worlds is
described in terms of two macroscopic parameters: $ x\equiv
{T'}/{T}, \quad \beta\equiv {\Omega'_B}/{\Omega_B}$. In Sec.XII we
describe the dark matter assuming that shadow baryons and shadow
helium atoms are the best candidates for DM. We explain the result
of astrophysical observations: $\Omega_{DM}/\Omega_{M}\simeq 5$.
In Sec.XIII we review the scenario of baryogenesis published in
our paper \ct{2}.

\section{Superstring theory and $E_6$ unification}

\subsection{Superstring theory}

Superstring theory \ct{21,23,26} is a paramount candidate for the
unification of all fundamental gauge interactions with gravity.
Superstrings are free of gravitational and Yang-Mills anomalies if
the gauge group of symmetry is $SO(32)$ or $E_8\times E_8$. The
'heterotic' superstring theory $E_8\times E'_8$ was suggested as a
more realistic model for unification of all fundamental gauge
interactions with gravity \ct{23}. However, this ten-dimensional
theory can undergo spontaneous compactification. The integration
over six compactified dimensions of the $E_8$ superstring theory
leads to the effective theory with the $E_6$ unification in the
four-dimensional space \ct{26}.

Superstring theory has led to the speculation that there may exist
another form of matter -- hidden ``shadow matter'' -- in the
Universe, which only interacts with ordinary matter via gravity or
gravitational-strength interactions \ct{29}. The shadow world, in
contrast to the mirror world \ct{27,28}, can be described by
another group of symmetry (or by a chain of groups of symmetry),
which is different from the ordinary world symmetry group.
According to the (super)string theory, the two worlds, ordinary
and shadow, can be viewed as parallel branes in a higher
dimensional space, where O-particles are localized on one brane
and H-particles - on another brane, and gravity propagates in the
bulk.

In our model we have assumed that at very high energies there
exists the $E_6$ unification predicted by superstring theory.

\subsection{$E_6$ Unification}

Three 27-plets of $E_6$ contain three families of quarks and
leptons, including right-handed neutrinos $N_a^c$ (where $a=1,2,3$
is the index of generations). The description of generations is
briefly discussed in Ref.~\ct{50}, but here we omit generation
subscripts, for simplification.

Matter fields (quarks, leptons and scalar fields) of the
fundamental 27-representation of the $E_6$ group decompose under
$SU(5)\times U(1)_X$ subgroup as follows (see Ref.~\ct{46}):
\be
        27 \to (10,1)  + (\bar 5, 2)+
          (5,-2)+ (\bar 5,-3)  + (1,5) + (1,0).    \lb{1a} \ee
The first and second numbers in the brackets in Eq. (\ref{1a})
correspond to the dimensions of the $SU(5)$ representations and to
the $U(1)_X$ charges, respectively. These representations
decompose under the groups with the breaking
\be SU(5)\times U(1)_X \to SU(3)_C\times SU(2)_L\times
U(1)_Z\times U(1)_X.            \lb{3a} \ee
We consider the following $U(1)_Z\times  U(1)_X$ charges of matter
fields: $Z=\sqrt{\frac{5}{3}}Q^Z,\,X=\sqrt{40}Q^X$.

The Standard Model (SM) family which contains the doublets of
left-handed quarks $Q$ and leptons $L$, right-handed up and down
quarks $u^c$, $d^c$, and also right-handed charged lepton $e^c$,
belongs to the $(10,1) + (\bar 5,2)$ representations of
$SU(5)\times U(1)_X$. Then, for the decomposition (\ref{3a}), we
have the following assignments of particles:
\bea
       (10,1) \to Q = &\left(\begin{array}{c}u\\
                                          d \end{array}\right) &\sim
                         \left(3,2,\frac 16,1\right),\nonumber\\
&u^{\rm\bf c} &\sim \left(\bar3,1,-\frac 23,1\right),\nonumber\\
&e^{\rm\bf c} &\sim \left(1,1,1,1\right).       \lb{4a}\\
(\bar 5,2) \to &d^{\rm\bf c}&\sim \left(\bar 3,1,\frac
13,2\right),
\nonumber\\
L = &\left(\begin{array}{c}e\\
                                             \nu \end{array}\right) &\sim
                         \left(1,2,-\frac 12,2\right),               \lb{5a}\\
(1,5) \to & S\,\,\, &\sim \,\,\left(1,1,0,5\right).\lb{6a} \eea
The remaining representations in (\ref{3a}) decompose as follows:
\bea
        (5,-2) \to& D&\sim \left(3,1,-\frac 13,-2\right),\nonumber\\
                   h = &\left(\begin{array}{c}h^+\\
                                               h^0 \end{array}\right) &\sim
                         \left(1,2,\frac 12,-2\right).
                                                              \lb{7a}\\
    (\bar 5,-3) \to &D^{\rm\bf c} &\sim \left(\bar 3,1,\frac 13,-3\right),
\nonumber\\
                     h^{\rm\bf c} = &\left(\begin{array}{c}h^0\\
                                               h^- \end{array}\right) &\sim
                         \left(1,2,-\frac 12,-3\right).              \lb{8a}
\eea
To the representation (1,5) is assigned the SM-singlet field S,
which carries nonzero $U(1)_X$ charge. The light Higgs doublets
are accompanied by the heavy colour triplets of exotic quarks
('diquarks') $D,D^{\rm\bf c}$ which are absent in the SM (see
Ref.~\ct{46}).

The right-handed heavy neutrino is a singlet field $N^c$
represented by (1,0):
\be
       (1,0) \to N^{\rm\bf c} \sim  (1,1,0,0).             \lb{9a} \ee

\subsection{Breaking of the $E_6$ Unification}

It is well known (see, for example, Ref. \ct{55}) that there exist
three schemes of breaking the $E_6$ group:
\begin{eqnarray}
i)\,\,\, E_6 &\to& SU(3)_1\times SU(3)_2\times SU(3)_3,\label{break1}\\
ii)\,\,\, E_6&\to& SO(10)\times U(1),\label{break2}\\
iii)\,\,\, E_6&\to& SU(6)\times SU(2)\label{break3}.
\end{eqnarray}
The first case was considered in the first paper of Refs.~\ct{1},
where we have investigated the possibility of the breaking:
\be E_6\to SU(3)_C\times SU(3)_L\times SU(3)_R \lb{1BE} \ee
in both O- and M-worlds, with broken mirror parity. The model has
the merit of an attractive simplicity. However, in such a model we
are unable to explain the tiny CC (\ref{12AC})\, given by
astrophysical measurements, because in the case (\ref{1BE})\, we
have in both worlds the low-energy limit of the SM, which forbids
a large confinement radius (i.e. small energy scale) of any
interaction.

It is quite impossible to obtain the same $E_6$ unification in the
O- and M-worlds with the same breakings $ii)$ or $iii)$ in both
worlds if  mirror parity MP is broken. In this case, we are forced
to assume different breakings of the $E_6$ unification in
the O- and H-worlds:\\
$$E_6 \to SO(10)\times U(1) \qquad {\rm in\,\,\, O-world}, $$
$$ E'_6 \to SU(6)'\times SU(2)'\qquad {\rm in\,\,\, H-world}, $$
explaining the small value of the CC, $\Lambda$, by condensation
of fields belonging to the additional $SU(2)'$ gauge group which
exists only in the H-world and has a large confinement radius.

The breaking mechanism of the $E_6$ unification is given in
Ref.~\ct{56}. The vacuum expectation values (VEVs) of the Higgs
fields $H_{27}$ and $H_{351}$ belonging to 27- and 351-plets of
the $E_6$ group can appear in the case \eqref{break2} for the
O-world only with nonzero 27-component:
\be  \langle H_{351}\rangle=0, \qquad v=\langle H_{27}\rangle\neq
0. \lb{2BE} \ee
In the case \eqref{break3} for the H-world we have
\be  \langle H_{27}\rangle=0, \qquad V=\langle H_{351}\rangle\neq
0. \lb{3BE} \ee
The 27 representation of $E_6$ is decomposed into 1 + 16 + 10
under the $SO(10)$ subgroup and the 27 Higgs field $H_{27}$ is
expressed in 'vector' notation as
\bea
   H_{27}\equiv
    &\left(\begin{array}{c}H_0\\H_{\alpha}\\
                                          H_M\end{array}\right),&
\eea
where the subscripts 0, $\alpha=1,2,...,16$ and $M=1,2,...,10$
stand for  singlet, the 16- and the 10-representations of
$SO(10)$, respectively. Then
\bea
   \langle H_{27}\rangle=
    &\left(\begin{array}{c}v\\0\\
                                          0\end{array}\right).&
\eea
Taking into account that the 351-plet of $E_6$ is constructed from
$27\times 27$ symmetrically, we see that the trace part of
$H_{351}$ is a singlet under the maximal little groups. Therefore,
in a suitable basis, we can construct the VEV $\langle
H_{351}\rangle$  for the case of the maximal little group
$SU(2)\times SU(6)$. A singlet under this group which we get from
a symmetric product of $27\times 27$ comes from the component $(1,
15)\times (1, 15)$ and hence
\bea \langle H_{351}\rangle =&\left(\begin{array}{cc}V\otimes
1_{15}&\\&0\otimes 1_{15}\end{array}\right).& \eea
According to the assumptions of Ref.~\ct{1}, in the ordinary world
there exists the following chain of symmetry groups from the GUT
scale of the $E_6$ unification up to the Standard Model (SM)
scale:
$$
E_6\to SO(10)\times U(1)_Z \to SU(4)_C\times SU(2)_L \times
SU(2)_R\times U(1)_Z$$ $$ \to SU(3)_C\times SU(2)_L \times
SU(2)_R\times U(1)_X\times
 U(1)_Z $$ \be \to [SU(3)_C\times SU(2)_L\times U(1)_Y]_{{SUSY}}\to
 SU(3)_C\times SU(2)_L\times U(1)_Y. \lb{4BE}
 \ee
In the shadow H-world, we have the following chain:
$$ E'_6 \to SU(6)'\times SU(2)'_{\theta}
\to SU(4)'_C\times SU(2)'_L\times SU(2)'_{\theta}\times U(1)'_Z $$
$$\to SU(3)'_C\times
SU(2)'_L\times SU(2)'_{\theta}\times U(1)'_X \times U(1)'_Z$$ \be
\to [SU(3)'_C\times SU(2)'_L\times SU(2)'_{\theta}\times
U(1)'_Y]_{{SUSY}}\to SU(3)'_C\times SU(2)'_L\times
SU(2)'_{\theta}\times U(1)'_Y .    \lb{5BE} \ee
In general, this is not an unambiguous choice of the $E_6(E'_6)$
breaking chains.

\section{$E_6$ unification in ordinary and mirror world}

The results of Refs.~\ct{32,33,36,38,39,41,42} are based on the
hypothesis of the existence in Nature of a mirror (M) world
parallel to the visible ordinary (O) world. The authors have
described the O- and M-worlds at low energies by a minimal
symmetry $G_{SM}\times G'_{SM}$ where
$$G_{SM} = SU(3)_C\times SU(2)_L\times U(1)_Y$$ stands
for the observable Standard Model (SM) while
$$G'_{SM} = SU(3)'_C\times SU(2)'_L\times U(1)'_Y$$ is its mirror
gauge counterpart. The M-particles are singlets of $G_{SM}$ and
the O-particles are singlets of $G'_{SM}$. {\it These different O-
and M-worlds are coupled only by gravity, or possibly by another
very weak interaction}. In general, we can consider a
supersymmetric theory when $ G\times G'$ contains the grand
unification groups $SU(5)\times SU(5)'$, $SO(10)\times
SO(10)',\,\,$ $E_6\times E_6'$ etc.

\subsection{Particle content in the ordinary and mirror worlds}

The M-world is a mirror copy of the O-world and contains the same
particles and types of interactions as our visible world. The
observable elementary particles of our O-world have the
left-handed (V-A) weak interactions which violate P-parity. If a
hidden mirror M-world exists, then mirror particles participate in
the right-handed (V+A) weak interactions and have the opposite
chirality.

Lee and Yang were the first \ct{27} to suggest such a duplication
of the worlds, which restores the left-right symmetry of Nature.
They introduced a concept of right-handed particles, but their
R-world was not hidden. The term 'Mirror Matter' was introduced by
Kobzarev, Okun and Pomeranchuk \ct{28}. They suggested the 'Mirror
World' as the hidden sector of our Universe, which interacts with
the ordinary (visible) world only via gravity or another very weak
interaction. They have investigated a variety of phenomenological
implications of such parallel worlds (for recent comprehensive
reviews on mirror particles and mirror matter, see Refs. \ct{58}).

Including the Higgs bosons $\Phi$, we have the following SM
content of the O-world:
$$\rm{
L-set}: {\quad (u,d,e,\nu,\tilde u,\tilde d,\tilde e,\tilde
N)_L\,,\Phi_u,\,\Phi_d} ;$$
$$ \rm { \tilde R-set}: { \quad (\tilde
u,\tilde d,\tilde e,\tilde \nu,u,d,e,N)_R\,,\tilde \Phi_u,\,\tilde
\Phi_d;}$$
with the antiparticle fields: ${\tilde \Phi_ {u,d} =
\Phi^*_{u,d},}\,\,$ ${ \tilde \psi_R = C\gamma_0\psi_L^*\,\,}$ and
${\tilde \psi_L = C\gamma_0\psi_R^*.}$

Considering the minimal symmetry $G_{SM}\times G'_{SM}$, we have the
following particle content in the M-sector:
$$ \rm{ L'-set}: \quad \Large
(u',d',e',\nu',\tilde u',\tilde d',\tilde e',\tilde
N')_L\,,\Phi'_u,\,\Phi'_d ;$$
$${\rm
 \tilde R'-set}: {\quad (\tilde u',\tilde d',\tilde e',\tilde
\nu',u',d',e',N')_R\,,\tilde \Phi'_u,\,\tilde \Phi'_d.}$$

\subsection{Mirror world with broken mirror parity}

If the ordinary and mirror worlds are identical, then O- and
M-particles should have the same cosmological densities. But this
is immediately in conflict with recent astrophysical measurements.
Mirror parity (MP) is not conserved, and the ordinary and mirror
worlds are not identical. Then the VEVs of the Higgs doublets
$\phi$ and $\phi'$ are not equal:
\be \langle\phi\rangle=v,\quad \langle\phi'\rangle=v'\quad
{\rm{and}} \quad v\neq v'.  \lb{1MW} \ee
Introducing the parameter characterizing the violation of MP:
\be       \zeta = \frac{v'}{v} \gg 1, \lb{2MW}  \ee
we have the estimate of Refs.~\ct{32,33,36,38,39,41,42}:
$$\zeta \sim 100$$.

Then the masses of fermions and massive bosons in the mirror world
are scaled up by the factor $\zeta$ with respect to the masses of
their counterparts in the ordinary world:
\begin{eqnarray}
               m'_{q',l'} &=& \zeta m_{q,l}, \lb{3MW} \\
                 M'_{W',Z',\Phi'} &=& \zeta M_{W,Z,\Phi}, \lb{4MW}
                 \end{eqnarray}
while photons and gluons remain massless in both worlds.

Let us consider now the expressions for the running of the inverse
coupling constants,
\begin{eqnarray}
    \alpha_i^{-1}(\mu) &=& \frac{b_i}{2\pi}\ln
    \frac{\mu}{\Lambda_i},\ \quad \mbox{in the O-world;}  \lb{5MW}\\
{\alpha'}_i^{-1}(\mu) &=& \frac{b'_i}{2\pi}\ln
    \frac{\mu}{\Lambda'_i}, \ \quad \mbox{in the M-world}.       \lb{6MW}
    \end{eqnarray}
Here $i=1,2,3$ correspond to $U(1),\, SU(2)$ and $SU(3)$ groups of
the SM (or SM$'$). A big difference between the electroweak scales
$v$ and $v'$ will not cause the same difference between the scales
$\Lambda_i$ and $\Lambda'_i.$ Hence,
\be \Lambda'_i = \xi \Lambda_i, \lb{7MW} \ee where $\xi > 1$.

\subsection{Seesaw scale in the ordinary and mirror
worlds}

In the language of neutrino physics, the O-neutrinos
$\nu_e,\,\,\nu_{\mu},\,\,\nu_{\tau}$ are active neutrinos, while
the M-neutrinos $\nu'_e,\,\,\nu'_{\mu},\,\,\nu'_{\tau}$ are
sterile neutrinos. The model \ct{32,33,36,38,39,41,42} provides a
simple explanation of why sterile neutrinos could be light, and
could have significant mixing with the active neutrinos.

If MP is conserved ($\zeta = 1$), then the neutrinos of the two
sectors are strongly mixed. But it seems that the situation with the
present experimental and cosmological limits on the active-sterile
neutrino mixing do not confirm this result. If instead MP is
spontaneously broken, and $\zeta \gg 1$, then the active-sterile
mixing angles should be small:
\be \theta_{\nu\nu'}\sim \frac 1{\zeta}. \lb{1SS} \ee
As a result, we have the following relation between the masses of
the light left-handed neutrinos:
\be
  m'_{\nu}\approx \zeta^2 m_{\nu}. \lb{2SS} \ee

In the context of the SM, in addition to the fermions with
non-zero gauge charges, one introduces also the gauge singlets,
the so-called right-handed neutrinos $N_a$ with large Majorana
mass terms. According to Refs.~\ct{32,33,36,38,39,41,42}, they
have equal masses in the O- and M-worlds:
\be M'_{\nu,a} = M_{\nu,a}. \lb{3SS} \ee

Let us consider now the usual seesaw mechanism. Heavy right-handed
neutrinos are created at the seesaw scales $M_R$ in the O-world
and $M'_R$ in the M-, or H-world. From the Lagrangian, considering
the Yukawa couplings identical in the two sectors, it follows that
\be m_{\nu}^{(')}=\frac {{v^{(')}}^2}{M_R^{(')}}, \lb{4SS} \ee and
we immediately obtain the relations (\ref{2SS}), with
\be M'_R = M_R \lb{5SS}. \ee
Then we see that even in the model with broken mirror parity, we
have the same seesaw scales in the O- and M-(H-)worlds.

\section{Shadow world and theta-particles}

In the first paper of Refs.~\ct{1} we have presented an example of
the gauge coupling constant evolutions from the SM up to the $E_6$
unification scale in the ordinary and mirror worlds with broken
mirror parity, assuming that the $E_6$ group of symmetry undergoes
the breaking:
         $ E_6 \to SU(3)_C\times SU(3)_L\times SU(3)_R $
in both worlds (O and M) and gives the SM group of symmetry at
lower energies. Of course, such a Universe could exist, but it is
difficult to find a simple explanation why the observable CC has
such a tiny value (\ref{12AC}), since such a model does not have
an extremely large radius of confinement of any gauge interaction.
Thus, it is impossible to conceive a vacuum with extremely small
vacuum energy density.

In the present paper we consider the idea of the existence of
theta-particles, developed by Okun \ct{60}\footnote{We are
grateful to M. Yu. Khlopov for this information.}. In those works
it was suggested the hypothesis that in Nature there exists the
symmetry group
\be SU(3)_C\times SU(2)_L\times SU(2)_{\theta}\times U(1)_Y \,,
\lb{1SW} \ee
i.e. with an additional non-Abelian $SU(2)_{\theta}$ group whose
gauge fields are neutral, massless vector particles -- 'thetons'.
These 'thetons' have a macroscopic confinement radius
$1/\Lambda_{\theta}$. Later, in Refs. \cite{3}, it was assumed
that if any $SU(2)$ group with the scale $\Lambda_2 \sim 10^{-3}$
eV exists, then it is possible to explain the small value
(\ref{12AC}) of the observable CC. The latter idea was taken up in
Refs.~\ct{3}.

In the present context we assume the existence of low-energy
symmetry group (\ref{1SW}) in the shadow world, but not in the
ordinary world, as a natural consequence of different schemes of
the $E_6$-breaking in the O- and H-worlds. $\theta-$particles are
absent in the ordinary world, because their existence is in
disagreement with all experiments. However, they can exist in the
H-world:
\be G' = SU(3)'_C\times SU(2)'_L\times SU(2)'_{\theta}\times
U(1)'_Y \,, \lb{2SW} \ee
By analogy with the theory developed in \ct{60}, we consider
shadow thetons ${\Theta'}^i_{\mu\nu}$, $i=1,2,3$, which belong to
the adjoint representation of the group $SU(2)'_{\theta}$, three
generations of shadow theta-quarks $q'_{\theta}$ and shadow
leptons $l'_{\theta}$, and the necessary $\theta$-scalars
$\phi'_{\theta}$ for the corresponding breakings. Shadow thetons
have macroscopic confinement radius
 $1/\Lambda'_{\theta}$, and we assume that
\be \Lambda'_{\theta}\sim 10^{-3}\,\, {\mbox{eV}}. \lb{3SW}   \ee

Matter fields  of the fundamental 27-representation of the $E'_6$
group decompose under $SU(2)'_\theta\times SU(6)'$ subgroup as
follows: 27=(2,6)+(1,15), where
\bea
       (2,6) \to q' = &\left(\begin{array}{c}q'_{\theta,A}|_{I_{\theta}=+1/2}\\
                                          q'_{\theta,A}|_{I_{\theta}=-1/2} \end{array}\right).
                                          &\\
        (1,15) \to & D',D'^c&\\
                   &h' = \left(\begin{array}{c}h'^+\\
                                               h'^0
                                               \end{array}\right),
                                               &\\
                    &h'^{\rm\bf c} = \left(\begin{array}{c}h'^0\\
                                               h'^-
                                               \end{array}\right),&\\
&{q'}^c_a, {N'}^c, S'.&
 \lb{4SW} \eea
Here $A=1,...,6; a=1,2,3$ are color indices and $I_{\theta}$ is a
$\theta$-isospin; $\theta-$quarks are $q'_{\theta,A}$, while
quarks ${q'}^c_a$, right-handed neutrino ${N'}^c$ and scalar $S'$
are $SU(2)'_{\theta}$-singlets.

\section{Inflation, $\bf E_6$ unification and the problem of walls in the Universe}

The simplest model of inflation is based on the superpotential
\be W=\lambda \varphi(\Phi^2-\mu^2), \lb{1i} \ee
containing the inflaton field given by $\varphi$ and the Higgs
field $\Phi$, where $\lambda$ is a coupling constant of order 1
and $\mu$ is a dimensional parameter of the order of the GUT scale
(see, for example, \ct{dvali}). The supersymmetric vacuum is
located at $\varphi=0$, $\Phi=\mu$, while for the field values
$\Phi=0$, $|\varphi| > \mu$ the tree level potential has a flat
valley with the energy density $V=\lambda^2\mu^4$. When the
supersymmetry is broken by the non-vanishing F-term, the flat
direction is lifted by radiative corrections and the inflaton
potential acquires a slope appropriate for the slow roll
conditions.

This so-called hybrid inflation model leads to the choice of the
initial conditions \ct{36}. Namely, at the end of the Planck epoch
the singlet scalar field $\varphi$ should have an initial value
$\varphi=f\sim 10^{18}$ GeV ($E_6$-GUT scale), while the field
$\Phi$ must be zero with high accuracy over a region much larger
than the initial horizon size $\sim M_{Pl}$. In other words, the
initial field configuration should be located right on the bottom
of the inflaton valley and the energy density starts with
$V=\lambda^2\mu^4 \ll M_{Pl}^4$.

If $E_6'$ is the mirror counterpart of $E_6$, then we have $Z_2$
symmetry, i.e. a discrete group connected with the mirror parity.
In general, the spontaneous breaking of a discrete group leads to
phenomenologically unacceptable walls of huge energy per area (see
Fig.~1).

\begin{center}
\scalebox{1} 
{
\begin{pspicture}(0,-4.004841)(11.174835,3.9848409)
\psbezier[linewidth=0.04,shadow=true,shadowangle=-45.0,fillstyle=gradient,gradlines=2000,gradbegin=red,
gradend=blue,gradmidpoint=1.0,gradangle=44.999996](1.1399997,2.7989092)(0.67250633,1.9149126)
(3.2434227,-1.1982028)(4.14,-1.6410909)(5.0365767,-2.083979)(10.0,-2.1810908)(10.24,-1.2810909)
(10.48,-0.38109082)(7.4934225,2.4118967)(6.58,2.8189092)(5.6665773,3.2259216)(1.607493,3.6829057)
(1.1399997,2.7989092)
\psbezier[linewidth=0.04,shadow=true,shadowangle=-45.0,fillstyle=gradient,gradlines=2000,
gradbegin=magenta,gradmidpoint=1.0,gradangle=90.0](0.58131564,-2.1189015)(1.1626313,-2.9810908)
(5.4029098,-1.4745029)(6.28339,-0.9561969)(7.1638703,-0.43789086)(11.08,2.1189091)
(9.852628,3.041875)(8.625256,3.964841)(5.2210903,2.7774158)(4.2657266,2.2189093)
(3.3103628,1.6604025)(0.0,-1.2567121)(0.58131564,-2.1189015)
\psbezier[linewidth=0.08,linecolor=green,doubleline=true,doublesep=0.12,doublecolor=yellow]
(5.58,3.7989092)(5.66,2.918909)(6.229719,3.9476376)(6.1,2.9789093)(5.97028,2.0101807)
(5.446328,2.2041657)(5.12,1.2589092)(4.793671,0.31365272)(3.9852622,-2.6786351)(4.9799995,-2.7810907)
\usefont{T1}{ptm}{m}{n}
\rput(2.897656,2.7339091){\large \color{white}$H_{27}$ on O-brane}
\usefont{T1}{ptm}{m}{n}
\rput(7.957656,-1.3460908){\large \color{white}$H_{27}$ on
Sh-brane}
\usefont{T1}{ptm}{m}{n}
\rput(2.5176558,-1.4660908){\large $H_{351}$ on O-brane}
\usefont{T1}{ptm}{m}{n}
\rput(7.997656,2.4939091){\large $H_{351}$ on Sh-brane}
\usefont{T1}{ptm}{m}{n}
\rput(6.8185935,1.0389092){\large Wall}
\psline[linewidth=0.06cm,arrowsize=0.05291667cm
4.0,arrowlength=2.0,arrowinset=0.4]{->}(6.14,0.9389092)(5.2,0.39890918)
\usefont{T1}{ptm}{m}{n}
\rput(5.106875,-3.7210908){\Large Fig. 1}
\end{pspicture}
}
\end{center}

\setcounter{figure}{1}

Then we have the following properties for the energy densities of
radiation, DM, M and wall:
$$ \rho_r\varpropto \frac 1{a(t)^4}, \quad \rho_{M,DM}\varpropto
\frac 1{a(t)^3}, \quad \rho_{wall}\varpropto \frac 1{a(t)}, $$
where $a(t)$ is a scale factor with cosmic time $t$ in the
Friedmann-Lemaitre-Robertson-Walker (FLRW) metric describing our
Universe. For large Universe we have
$\rho_{wall}\gg\rho_{M,DM},\rho_r.$ In our case of the hidden
world, the shadow superpotential is:
\be W'=\lambda' \varphi'({\Phi'}^2-\mu'^2), \lb{2i} \ee
where $\Phi'=H_{351}$ and $\langle H_{351}\rangle=\mu'$. Then the
initial energy density in the H-world is $V'={\lambda'}^2{\mu'}^4
\ll M_{Pl}^4$. To avoid this phenomenologically unacceptable wall
dominance we cannot assume symmetry under $Z_2$ and thus $V=V'$ is
not automatic. Instead, it is necessary to assume the following
fine-tuning:
\be V=V': \quad \lambda^2\mu^4={\lambda'}^2{\mu'}^4, \lb{3i} \ee
which helps to obtain the initial conditions for the GUT-scales
and GUT-coupling constants: $M_{E6}=M'_{E6'}$ and
$g_{E6}=g'_{E6'}$.

\section{The cosmological constant problem}

The cosmological constant (CC) was first introduced by Einstein in
1917 \ct{1CC} with aim to admit a static cosmological solution in
his new general theory of relativity. The introduction of CC
$\lambda$, the bare cosmological constant, was presented only by
the addition to the original field equations:
\be G^{\mu\nu}= R^{\mu\nu} - \frac{1}{2}Rg^{\mu\nu}= 8\pi
GT^{\mu\nu}     \lb{1CC} \ee
of the divergence-free term $-\lambda g_{\mu\nu}$:
\be G^{\mu\nu}= 8\pi GT^{\mu\nu} -\lambda g^{\mu\nu},    \lb{2CC}
\ee
where $R_{\mu\nu}$ is the Ricci curvature of $g_{\mu\nu}$, and
$T_{\mu\nu}$ is the energy-momentum tensor of matter.

Later it was realized (see \ct{2CC,3CC}) that quantum fluctuations
result in a vacuum energy, $\rho_{vac}$: any mode contributes
$\frac{1}{2}\hslash \omega$ to the vacuum energy, and the expected
value of the energy momentum tensor of matter is:
\be  \langle T^{\mu\nu}\rangle = T^{\mu\nu}_m -
\rho_{vac}g^{\mu\nu}, \lb{3CC} \ee
where $T^{\mu\nu}_m$ vanishes in vacuum. The quantum expectation
of the energy-momentum tensor, $\langle T^{\mu\nu}\rangle$, acts
as a source for the Einstein tensor, and we have:
\be G^{\mu\nu}= 8\pi GT^{\mu\nu}_m -\Lambda g^{\mu\nu}, \lb{4CC}
\ee
where $\Lambda$ is the effective cosmological constant provided by
the contribution of the vacuum energy, $\rho_{vac}$. We would
expect that the effective vacuum energy:
\be \rho_{vac}^{(eff)} = \frac{\lambda}{8\pi G} + \rho_{vac} =
\frac{\Lambda}{8\pi G} \lb{5CC} \ee
to be no smaller than $\rho_{vac}$. Even if the 'bare'
cosmological constant is assumed to vanish ($\lambda=0$), the
effective cosmological constant is not equal to zero. Requiring
that $\Lambda = 0$ means that there must be an exact cancellation
between the 'bare' cosmological constant, $\lambda$, and the
vacuum energy stress, $8\pi G\rho_{vac}$:
\be \Lambda=0 \quad \to \quad  \lambda + 8\pi G\rho_{vac}=0.
\lb{6CC} \ee

When the spontaneous symmetry breaking was widely discussed in the
Standard Model, Veltman commented that the vacuum energy arising
in spontaneous symmetry breaking gives an additional contribution
to the CC \ct{4CC}.

If we assume that the field theory is only valid up to some energy
scale $M_{cutoff}$, then there is a contribution to $\rho_{vac}$
of $O(M_{cutoff}^4)$. Collider experiments have established that
the SM is accurate up to energy scales $M_{cutoff}\gtrsim
O(M_{EW})$, where $M_{EW}\approx 246$ GeV is the electroweak
scale. We would therefore expect $\rho_{vac}$ to be at least
$O(M_{EW}^4)$.

In the absence of any new physics between the electroweak and the
Planck scale, $M_{Pl} \approx 1.2\times 10^{19}$ GeV, where
quantum fluctuations in the gravitational field can no longer be
safely neglected, we would expect $\rho_{vac}\sim O(M^4_{Pl})$. If
supersymmetry were an unbroken symmetry of Nature, the quantum
contributions to the vacuum energy would all exactly cancel
leaving $\rho_{vac}=0$ and $\Lambda=\lambda$. However, our
universe is not supersymmetric today, and so SUSY must have been
broken at some energy scale $M_{SUSY}$, where $1\,\,
{\rm{TeV}}\lesssim M_{\rm{SUSY}}\lesssim M_{Pl}$. It is necessary
to comment that the SUSY breaking is necessary in our superstring
and thereby SUSY-based model. We would expect $\rho_{vac}\sim
O(M^4_{\rm{SUSY}})$. Our model of quantum cosmology also had to
take into account extra dimensions and branes, spontaneous
breaking of compactification.

Previously in Ref.~\ct{1MPP} and also in Refs. \ct{2MPP} it was
shown that SUGRA models which ensure the vanishing of the vacuum
energy density near the physical vacuum lead to a natural
realization of the Multiple Point Model (MPP) \ct{4MPP} (see also
the reviews \ct{7MPP}) describing the degenerate vacua with zero
$\Lambda$.

The expansion rate of our Universe is sensitive to
$\rho_{vac}^{(eff)}$, or equivalently $\Lambda$. The result of
astrophysical measurements is given by Eq.~(\ref{12AC}), which
has established that ${(\rho_{vac}^{(eff)}})^{1/4}\simeq 2.3\times
10^{-3}$ eV. This implies that $\rho_{vac}^{(eff)}$ is some
$10^{60} - 10^{120}$ times smaller than the expected contribution
from quantum fluctuations, and gives rise to the cosmological
constant problem: {\it "Why is the measured effective vacuum
energy or cosmological constant so much smaller than the expected
contributions to it from quantum fluctuations?"}.

\section{A proposal for solving the CC problem}

Here we follow the ideas of Ref.~\ct{5CC}, which gives a possible
way to solve the CC problem.

In quantum mechanics we consider the probability amplitudes: The
initial state $|I\rangle$ transforming to a final state
$|F\rangle$. In this spirit, using the Euclidian action $S_E$,
only with the Ricci scalar $R$ and CC $\Lambda$, E.~Baum  and
S.~Hawking \ct{8CC} have calculated the path integral in the
Euclidian space-time which gives the following expression:
\be
      e^{-I_E} = e^{3\pi M_{Pl}/\Lambda}.  \lb{7CC} \ee
So, $\Lambda = 0$ dominates the action integral, which is
interpreted as the probability for $\Lambda = 0$ is close to 1.

The essence of the new approach \ct{5CC} is that the bare
cosmological constant $\lambda$, considered in Section VI, is
promoted from a parameter to a field. The minimization of the
action with respect to $\lambda$ then yields an additional field
equation, which determines the value of the effective CC,
$\Lambda$. In the classical history it dominates the partition
(wave) function of the Universe, $Z$.

If we take the total action of the Universe defined on a manifold
$\cal M$, and with effective cosmological constant $\Lambda$, to
be $I_{tot}(g_{\mu\nu}, \Psi^a, \Lambda;{\cal M})$, where $\Psi^a$
are the matter fields and $g_{\mu\nu}$ is the metric field, then
we define $I_{class}(\Lambda;{\cal M})$ to be the value of
$I_{tot}(g_{\mu\nu},\Psi^a,\Lambda;{\cal M})$ evaluated with
$g_{\mu\nu}$  and  $\Psi^a$  obeying their classical field
equations for fixed boundary initial conditions, and obtain the
field equation for the effective CC, $\Lambda$, given by
\be
   \frac{dI_{class}(\Lambda;{\cal M})}{d\Lambda} = 0.
 \lb{8CC} \ee
With a given $\cal M$, the equation for $\lambda$,
Eq.~(\ref{8CC}), can be viewed as a consistency equation which
relates the configuration of metric and matter variables in $\cal
M$ to $\lambda$. Eq.~(\ref{8CC}) can be viewed as a consistency
condition on the configuration of the effective CC, $\Lambda$, the
matter, $\Psi^a$, and metric, $g_{\mu\nu}$, fields in $\cal M$.
The consistency condition provided by Eq.~(\ref{8CC}) will be
violated for the vast majority of potential configurations
$\{g_{\mu\nu}, \Psi^a,\Lambda\}$. If observations determine a set
of $\{g_{\mu\nu}, \Psi^a,\Lambda\}$ for which Eq.~(\ref{8CC}) is
violated then this proposal would be falsified. At the same time,
if the observed configuration is consistent with Eq.~(\ref{8CC})
to within observational limits, then the present proposal would,
for the time being, have passed an important empirical test and
remain a credible solution to the CC problems. If $\Lambda \approx
0$ dominates the action integral, then we have an approximate
cancellation between the 'bare' cosmological constant and the
vacuum energy stress:
\be \Lambda \approx 0 \quad \to \quad  \lambda  \approx - 8\pi
G\rho_{vac}. \lb{9CC} \ee
The proposal \ct{5CC} for solving the CC is similar in certain
respects to other multiverse models such as the string landscape,
when $\Lambda$ takes different values in different vacua parts of
the multiverse. Despite this similarity, this proposal differs
from multiverse /landscape models. Also it is agnostic about the
modified theory of gravity and the number of space-time
dimensions.

\section{Dark energy}

\subsection{Quintessence model of cosmology}

Quintessence is described by a complex scalar field $\varphi$
minimally coupled to gravity. In the context of the General
Relativity (GR), the gravity is universal force described by the
space-time metric $g_{\mu\nu}$, and the dynamics of two worlds,
ordinary and hidden, is governed by the following action:
\be S= \int d^4x \sqrt{-g} \left[\frac {1}{2\kappa^2}R + \lambda +
{(\nabla \varphi)}^2 - V(\varphi) + L + L' + L_{mix}\right],
\lb{1g} \ee
where
\be {(\nabla \varphi)}^2 = g^{\mu\nu}\partial_{\mu}\varphi
\partial_{\nu}\varphi,   \lb{2g} \ee
and $V(\varphi)$ is the potential of the field $\varphi$,
$\kappa^2=8\pi G=M_{Pl}^{-2}$, $M_{Pl}$ is the reduced Planck
mass, $R$ is the space-time curvature, $\lambda$ is 'bare'
cosmological constant, $L(L')$ is the Lagrangian of the O-(H-)
sector, and $L_{mix}$ is the Lagrangian of photon-photon$'$,
neutrino-neutrino$'$, etc. mixing (see \ct{41}).

When both $E_6$ and $E'_6$ symmetry groups are broken, at the same
seesaw scales $M_R=M'_R$, down to $G_{SM}$ and $G'_{SM}\times
SU(2)'_\theta$ subgroups, respectively, then we have:
\be L = L_{gauge} +  L_{Higgs} +  L_{Yuk},\quad L' =  L'_{\theta}
+ L'_{gauge} +  L'_{Higgs} +  L'_{Yuk}, \lb{3g} \ee
where all parts of Lagrangians $L$ and $L'$ are self-explanatory.

The two sectors mean that at least below the scales $M_R=M'_R$ the
degrees of freedoms (the fields) can be classified into fields
from section O and fields from section H. We could thus consider
the energy density due to zero point fluctuations in the H-fields
as contributing to $\rho^{(H)}_{vac}$ while the O-fields
contribute to $\rho^{(O)}_{vac}$. Here we see that
\be \rho^{(O)}_{vac}= \rho^{(SM)}_{vac},
 \lb{4g} \ee
and
\be \rho^{(H)}_{vac}= \rho^{(SM')}_{vac} + \rho^{(\theta)}_{vac}.
 \lb{5g} \ee
Taking into account the fine-tuning considered in Section V, we
can assume that the SUSY breaking scales are identical in O- and
H-worlds: $M_{SUSY}=M'_{SUSY}$. Then
\be \rho^{(SM)}_{vac} = \rho^{(SM')}_{vac}\sim O(M^4_{\rm{SUSY}}),
 \lb{6g} \ee
and
\be \rho^{(H)}_{vac}= \rho^{(O)}_{vac} + \rho^{(\theta)}_{vac}.
 \lb{7g} \ee
In the framework of our cosmological model we calculate the dark
energy density relating the value $\rho_{DE}$ only with the
$SU(2)'_{\theta}$ gauge group contributions. This explains the
smallness of the dark energy density given by astrophysical
measurements. This phenomenon is not obvious, but we can try to
explain it.

If we neglect the weak connection between O- and H-worlds via
gravity, then we can approximately consider them as independently
existing in the Universe, and each sectors can be described by
their own actions with 'bare' cosmological constant $\lambda_0$:
\be S_O = \int d^4x \sqrt{-g} \left[\frac {1}{2\kappa^2}R +
\lambda_0 + {(\nabla \varphi)}^2 - V(\varphi) + L \right], \lb{8g}
\ee
and
\be S_H = \int d^4x \sqrt{-g} \left[\frac {1}{2\kappa^2}R +
\lambda_0 + {(\nabla \varphi')}^2 - V(\varphi') + L' +
L'_{\theta}\right]. \lb{9g} \ee
According to the proposal \ct{5CC}, the most probable is the
extremum given by Eq.~(\ref{8CC}) for the ordinary world:
\be  \lambda_0 + 8\pi G\rho^{(O)}_{vac}=0. \lb{10g} \ee
Then
\be \rho^{(O,eff)}_{vac}=0,
 \lb{11g} \ee
and
\be \rho^{(H,eff)}_{vac}= \rho^{(\theta)}_{vac}.
 \lb{12g} \ee
Finally, we obtain:
\be  \rho^{(eff)}_{vac} = \rho^{(H,eff)}_{vac} =
\rho^{(\theta)}_{vac}, \lb{13g} \ee
Here the effective CC, $\Lambda$, is not zero:
 \be \Lambda = 8\pi
G\rho^{(\theta)}_{vac}, \lb{14g} \ee
and the effective vacuum energy density is equal to DE density:
\be   \rho_{DE} = \rho^{(eff)}_{vac} = \rho^{(\theta)}_{vac}.
\lb{15g} \ee
This speculative consideration explains a tiny value of the DE
density calculated into the next Subsection.

\subsection{Inflaton, axion and DE density}

We assume that there exists an axial $U(1)_A$ global symmetry in
our theory, which is spontaneously broken at the scale $f$ by a
singlet complex scalar field $\varphi$:
\be \varphi = (f + \sigma) \exp(ia_{ax}/f). \lb{1q} \ee
We assume that a VEV $\langle \varphi \rangle = f$ is of the order
of the $E_6$ unification scale: $f\sim 10^{18}$ GeV. The real part
$\sigma $ of the field $\varphi$ is the inflaton, while the boson
$a_{ax}$ (imaginary part of the singlet scalar fields $\varphi$)
is an axion and could be identified with the massless
Nambu-Goldstone (NG) boson if the corresponding $U(1)_A$ symmetry
is not explicitly broken by the gauge anomaly. However, in the
hidden world the explicit breaking of the global $U(1)_A$ by
$SU(2)'_\theta$ instantons inverts $a_{ax}$ into a pseudo
Nambu-Goldstone (PNG) boson $a_{\theta}$. Therefore, in the
H-world we have:
\be \varphi' = (f + \sigma') \exp(ia_{\theta}/f). \lb{2q} \ee
The flat FLRW spacetime gives the following field equation for
axion $a_{\theta}$ (see reviews \ct{18}):
\be \frac{d^2 a_\theta}{dt^2} + 3 H \frac{d a_\theta}{dt} +
V'(a_\theta) = 0. \lb{3q} \ee
where $H$ is the Hubble parameter.

The singlet complex scalar field $\varphi$ reproduces a
Peccei-Quinn (PQ) model \ct{14CC}.  Near the vacuum, a PNG mode
$a_\theta$ emerges the following PQ axion potential:
\be V_{PQ}(a_\theta) \approx {(\Lambda'_{\theta})}^4
         \left(1 - \cos(a_\theta/f)\right).   \lb{4q} \ee
This axion potential exhibits minima at
\be {(V_{PQ})}|_{min} = 0,  \lb{5q} \ee
where:
\be \cos(a_\theta/f) = 1,\quad \rm{i.e.}\quad {(a_\theta)}_{min}=
2\pi n f, \quad n = 0,1,... \lb{6q} \ee
For small fields $a_\theta$ we expand the effective PQ potential
near the minimum:
\be V_{PQ}(a_\theta) \approx \frac{({\Lambda'_{\theta})}^4}{2f^2}
(a_\theta)^2 + ... = \frac 12 m^2 {(a_\theta)}^2 + ..., \lb{7q}
\ee
and hence the PNG axion mass squared is given by:
\be m^2\sim {(\Lambda'_{\theta})}^4/f^2.  \lb{8q} \ee
Solving Eq.~(\ref{3q}) for  $a_\theta$ we can use the axion
potential:
\be V(a_\theta)=V_{PQ}(a_\theta), \lb{9q} \ee
which gives:
\be V'(a_\theta)
=\frac{{(\Lambda'_\theta)}^4}{f}\sin(a_{\theta}/f). \lb{10q} \ee
If now $\sin(a_\theta/f)=0$, then ${\dot a}_\theta =0$, and
$V_{PQ}(a_\theta)=0$, because $\cos(a_\theta/f)=1$, according to
Eqs.~(\ref{4q}) and (\ref{5q}).

The minimum of the total $\theta$-potential is:
\be V_{\theta}|_{min} = V_{PQ}(a_\theta)|_{min} +
V_{\theta-condensate}, \lb{11q} \ee
where the first term is zero, according to Eq.(\ref{5q}), and
\be V_{\theta-condensate} = {(\Lambda'_\theta)}^4. \lb{12q} \ee
In this case when $a_\theta={\rm{const}}$ and ${\dot a}_\theta =0
$, the contribution of axions to the energy density of the
H-sector is equal to zero. Finally, we obtain:
\be  \rho^{(eff)}_{vac} = \rho^{(\theta)}_{vac} = |{\dot
a}_\theta|^2 + V_{\theta}|_{min} = {(\Lambda'_\theta)}^4. \lb{13q}
\ee
The DE density is equal to the value:
\be \rho_{DE} =  \rho^{(eff)}_{vac} = {(\Lambda'_\theta)}^4.
\lb{14q} \ee
Taking into account the result (\ref{12AC}) of recent
astrophysical observations, we obtain the estimate of the
$SU(2)'_\theta$ group's gauge scale:
\be \Lambda'_\theta \backsimeq  2.3\times 10^{-3}\,\,\rm{eV}.
\lb{16q} \ee
If $\Lambda'_\theta \sim 10^{-3}$ eV and $f\sim 10^{18}$ GeV,  we
can estimate the $\theta$-axion mass from Eq.~(\ref{8q}):
\be m\sim {\Lambda'_{\theta}}^2/f\sim 10^{-42} \,\,{\rm{GeV}},
\lb{17q} \ee
which is extremely small. But according to
Eqs.~(\ref{11q})-(\ref{14q}), these light axions do not give the
contribution to $\rho_{DE}$. It is given only by the condensate of
$\theta$-fields.

Then it is well-known (see reviews \ct{18}) that the equation of
state for $\theta$-fields is:
\be w_\theta = \frac{\dot{a}_{\theta}^2 -
2V_{\theta}}{\dot{a}_{\theta}^2 + 2V_{\theta}}, \lb{18q} \ee
and we have (with ${\dot a}_\theta =0$):
\be w=w_\theta=-1, \lb{19q} \ee
in accordance with the astrophysical observation (\ref{11AC}).

\section{Inflation in the ordinary and shadow worlds}

The results of Wilkinson Microwave Anisotropy Probe (WMAP) \ct{9}
lead to the severe constraint on inflationary models giving the
value of the spectral index:
\be    n_s = 0.95 \pm 0.02.   \lb{1i} \ee
The modern inflationary models give an exact scale-invariant
spectrum with $n_s=1$ (see \ct{14,61}). By this reason, any model
describing the early inflationary era has to take into account
this constraint: the inflationary potential, describing the early
inflationary universe, has to give the desired spectral index
$n_s$.

The scalar field $\varphi$ produces the following Coleman-Weinberg
potential \ct{62}:
\be V_{CW}= A{(\varphi^+\varphi)}^2\left (\log
{(\frac{\varphi^+\varphi}{f^2})}^2 - 1\right) + Af^4.   \lb{5i}
\ee
Then for the inflaton $\sigma^{(')}$ we can consider the following
inflationary potential in the zero temperature limit \ct{63}:
 \be V^{(')}_{infl}=
A^{(')}{(\sigma^{(')} + f)}^4\left (\log {(\frac{\sigma^{(')} +
f}{f})}^4 - 1\right) + A^{(')}f^4. \lb{6i} \ee
Taking into account the finite temperature effects, we have:
 \be V^{(')}_{infl, T^{(')}}= V_{infl}^{(')} + \beta_T^{(')} {(T^{(')})}^2(\sigma^{(')} + f)^2,
\lb{7i} \ee
where $\beta_T^{(')}$ is a constant. For compactness of notation,
here and in the following we denote the ordinary world rates by
the non-primed symbols and mirror-hidden world ones by the primed
symbols.

At high temperature, the field $\sigma$ is trapped at the $U(1)_A$
symmetric minimum $\langle\sigma\rangle = -f$ (i.e.
$\langle\varphi\rangle=0$). When the universe cools and gets a
sufficiently low temperature, then a new minimum appears at the
$U(1)_A$-symmetry breaking value $\langle\sigma\rangle=0$ (i.e.
$\langle\varphi\rangle=f$). The critical temperature $T_{cr}$
corresponds to such a value of temperature when the two above
minima become degenerate:
\be T_{cr}=f\sqrt{\frac{A}{\beta_T}}e^{-\frac 14}.\lb{8i} \ee
Then the Universe cools further and reaches the Hawking
temperature:
\be  T_{hawk} = \frac{H}{2\pi}\approx \frac{1}{2\pi}\sqrt
{\frac{8\pi}{3M_{Pl}^2} V_{infl}|_{\sigma=-f}} =
\sqrt\frac{A}{3\pi} \frac{f^2}{M_{Pl}}, \lb{9i} \ee
where $H$ is the Hubble parameter at that epoch. The first order
phase transition occurs and $\sigma$ starts its slow-rolling
towards the true minimum of the inflationary potential and gets
this minimum at the end of inflation. We have a similar
development in the hidden sector of the Universe.

But these two sectors, ordinary and hidden, have different
cosmological evolutions. In particular, they never had to be in
equilibrium with each other: the Big Bang Nucleosynthesis (BBN)
constraints require that H-sector must have smaller temperature
than O-sector: $T'<T$ (see Ref.~ \ct{42}).

\section{Reheating and radiation}

During reheating the exponential expansion, which was developed by
inflation, ceases and the potential energy of the inflaton field
decays into a hot relativistic plasma of particles. At this point,
the Universe is dominated by radiation, and then quarks and
leptons are formed.

All the difference between the ordinary and shadow worlds can be
described in terms of two macroscopic (free) parameters of the
model:
\be x\equiv \frac{T'}{T}, \quad \beta\equiv
\frac{\Omega'_B}{\Omega_B}, \lb{1r} \ee
where $T(T')$ is O-(H-) photon temperature in the present
Universe, and $\Omega_B(\Omega'_B)$ is O-(H-)baryon fraction.

In Subsection I.A we have presented  energy density ratio which is
a sum of relativistic (radiation) component $\Omega_r$,
non-relativistic (matter) component $\Omega_m$ and the vacuum
energy density $\Omega_{\Lambda}$. The modern observational data
indicate that the Universe is almost flat giving Eq.~(\ref{3AC}),
in a perfect accordance with the inflationary paradigm.

The relativistic fraction is represented by photons and neutrinos.
The contribution of the H-degrees of freedom to the observable
Hubble expansion rate, which are equivalent to an effective number
of extra neutrinos $\Delta N_{\nu}=6.14\cdot x^4$, is small
enough: $\Delta N_{\nu}=0.05$ for $x=0.3$ (see \ct{42}). In our
model:
\be \omega_r=\Omega_r h^2=4.2\cdot 10^{-5}(1+x^4),\quad
h=\frac{H}{H_0}, \lb{2r} \ee
where the contribution of H-species is negligible due to the BBN
constraint: $x^4\ll 1$.

Recent cosmological observations \ct{14} show that for redshifts
$(1+z) \gg 1$ we have:
\be    H(z) = H_0[\Omega_r(1+z)^4 + \Omega_m(1+z)^3]. \lb{3r} \ee
Therefore, the radiation is dominant at the early epochs of the
Universe, but it is negligible at present epoch:
$\Omega_r^{(0)}\ll 1$.

Any inflationary model have to describe how the SM-particles were
generated at the end of inflation. The inflaton, which is a
singlet of $E_6$, can decay, and the subsequent thermalization of
the decay products can generate the SM-particles. The inflaton
$\sigma$ produces gauge bosons: photons, gluons, $W^{\pm}$, $Z$,
and matter fields: quarks, leptons and the Higgs bosons, while the
inflaton field $\sigma'$ produces H-world particles: shadow
photons and gluons, thetons, $W'$, $Z'$, theta-quarks
$q_{\theta}$, theta-leptons $l_{\theta}$, shadow quarks $q'$ and
leptons $l'$, scalar bosons $\phi_{\theta}$ and shadow Higgs
fields $\phi'$. In shadow world we end up with a thermal bath of
$SM'$ and $\theta$ particles. However, we assume that the density
of $\theta$ particles is not too essential in cosmological
evolution due to small $\theta$ coupling constants.

According to Ref.~\ct{42}, at the end of inflation the O- and
H-sectors are reheated in a non-symmetric way ($T_R>T'_R$). After
reheating (at $T<T_R$) the exchange processes between O- and
H-worlds are too slow, by reason of very weak interaction between
two sectors. As a result, it is impossible to establish
equilibrium between them. So that both worlds evolve adiabatically
and the temperature asymmetry ($T'/T < 1$) is approximately
constant in all epochs from the end of inflation until the present
epoch. Therefore, the cosmology of the early H-world is very
different from the ordinary one when we consider such crucial
epochs as baryogenesis and nucleosynthesis. Any of these epochs is
related to an instant when the rate of the relevant particle
process, $\Gamma(T)$, becomes equal to the Hubble expansion rate
$H(T)$. In the H-world these events take place earlier and the
processes freeze out at larger $T$ than in the ordinary world.

\section{Big Bang Nucleosynthesis (BBN)}

At the end of cosmic inflation the Universe was filled with a
quark-gluon plasma. This plasma cools until  {\it the hadron
epoch} when hadrons (including baryons) can form. Then neutrinos
decouple and begin travelling freely through space. This cosmic
neutrino background is analogous to the CMB which was emitted much
later. After hadron epoch the majority of hadrons and anti-hadrons
annihilate each other, leaving leptons and anti-leptons dominating
the mass of the Universe. Here we reach {\it the lepton epoch}.
Then the temperature of the Universe continues to fall and falls
until the stop of the lepton/anti-lepton pairs creation. Also the
most leptons/anti-leptons are eliminated by annihilation
processes. At the end of the lepton epoch the Universe undergoes
{\it the photon epoch} when the energy of the Universe is
dominated by photons, which still essentially interact with
charged protons, electrons and eventually nuclei.

The temperature of the Universe again continues to fall. It falls
to the point when atomic nuclei begin to form. Protons and
neutrons combine into atomic nuclei by nuclear fusion process.
However, this nucleosynthesis stops at the end of the nuclear
fusion. At this time, the densities of non-relativistic matter
(atomic nuclei) and relativistic radiation (photons) are equal.

The BBN epoch in the H-world proceeds differently from ordinary
one and predicts different abundances of primordial elements. This
shadow BBN is analogous to the mirror BBN scenario considered in
Refs.~\ct{39,41,42}.

The difference of the temperatures ($T'<T$) gives that the number
density of H-photons is much smaller than for O-photons:
\be  \frac {n'_{\gamma}}{n_{\gamma}}=x^3\ll 1. \lb{4r} \ee
The primordial abundances of light elements depend on the baryon
to photon number density ratio: $\eta=n_B/n_{\gamma}$. The result
of WMAP \ct{9} gives: $\eta\simeq 6\cdot 10^{-10}$, in accordance
with the observational data.

The universe expansion rate at the ordinary BBN epoch (with $T\sim
1$ MeV) is determined by the O-matter density itself. As far as
$T'\ll T$, for the ordinary observer it is difficult to detect the
contribution of H-sector, which is equivalent to $\Delta
N_{\nu}\approx 6.14x^4$ and negligible for $x\ll 1$
 \ct{42}. As for the BBN epoch in the shadow world, for the H-observer the
contribution of O-sector is equivalent to $\Delta N'_{\nu}\approx
6.14x^{-4}$, which is dramatically large. Therefore, the observer
in H-world, which measures the abundances of shadow light
elements, should immediately detect the discrepancy between the
universe expansion rate and H-matter density at the shadow BBN
epoch (with $T'\sim 1 $ MeV): the O-matter density is invisible
for the H-observer.

During the structure formation, the most important moments are
connected with the matter-radiation equality (MRE), plasma
recombination and matter-radiation decoupling (MRD) epochs.

>From Eq.~(\ref{3r}) we see that MRE is given by the following
relation:
\be 1+z_{eq} = \frac{\Omega_m}{\Omega_r}. \lb{5r} \ee
The estimate of Ref.~\ct{39} gives:
\be 1+z_{eq} = 2.4\cdot 10^4\frac{\omega_m}{1+x^4}, \lb{6r} \ee
where $\omega_m=\Omega_m h^2$. The shadow relativistic component
is negligible for $x\ll 1$.

\subsection{Recombination}

The MRD takes place when the most of electrons and protons
recombine into neutral hydrogen and free electron density strongly
diminishes. During the recombination the photon scattering rate
drops below the Hubble expansion rate. In the O-world the MRD
takes place in the matter dominant period at the temperature
$T_{dec}\simeq 0.26$ eV corresponding to redshift:
\be 1+z_{dec} = \frac{T_{dec}}{T_{today}}\simeq 1100. \lb{7r} \ee
In the H-world we have the MRD temperature $T'_{dec}\simeq
T_{dec}$ and
\be 1+z'_{dec}\simeq x^{-1}(1+z_{dec})\simeq \frac{1100}{x}.
                                        \lb{8r} \ee
This means that in the H-world MRD occurs earlier than in the
O-world. According to Ref.~\ct{39},
\be        x_{dec}=\frac{1+z_{dec}}{1+z_{eq}}\simeq
\frac{4.59\cdot 10^{-2}}{\omega_m}, \lb{9r} \ee
and H-photon decoupling epoch coincides with the MRE epoch.
Eq.~(\ref{9r}) gives a critical value of temperature, which plays
a very important role in cosmology: for $x<x_{eq}$ the H-photons
would decouple yet during the radiation dominated period.

Thus, at the end of recombination, most of the atoms in the
Universe is neutral, photons travel freely and the Universe
becomes transparent. The observable CMB is a picture of the
Universe at the end of this epoch.

\section{Baryon density and dark matter}

Shadow baryons (and shadow helium), which are invisible by
ordinary photons, are the best candidates for dark matter (DM).

Here we give an approximate estimate of baryon masses in the O-
and H-worlds. The most part of mass of nucleons (proton and
neutron) is provided with dynamical (constituent) quark masses
$m_q$ forming the nucleon. The dynamical quark mass is
\be  m_q\simeq m_0 + \Lambda_{QCD}, \lb{10r} \ee
where $m_0\sim 10$ MeV is a current mass of light quarks
$u,\,\,d$, and $\Lambda_{QCD}\simeq 300$ MeV. Then the nucleon
mass $M_B$ can be estimated as
\be   M_B\simeq 3m_q\simeq 1\,\, {\rm{GeV}}. \lb{11r} \ee
As to shadow current quark mass $m'_0$ (see Subsection III.B), we
have
\be  m'_0\simeq \zeta m_0 \sim 1\,\, {\rm{GeV}} \lb{12r} \ee
for $\zeta\sim 100$. This estimate gives the shadow nucleon mass
$M'_B$ equal to
\be  M'_B\simeq 3(m'_0 + \Lambda'_{QCD}). \lb{13r} \ee
Taking into account Eq.~(\ref{7MW}) and the estimate $\xi\simeq
1.5$ given by Ref.~\ct{41} (see also Ref.~\ct{1}), we obtain
$\Lambda'_{QCD}\simeq 450$ MeV,  and:
\be  M'_B\simeq 3(1+0.45)\,\, {\rm{GeV}}\simeq 4.35\,\,{\rm{GeV}}.
                                                \lb{14r} \ee
Here we want to comment that in our model baryons of shadow world
are formed not only by quark system $qqq$, but also by
$q_{\theta,\vartheta}q_{\theta}^{\vartheta}q$, where
$\vartheta=1,2$ is the index of $SU(2)'_\theta$-group. The last
system gives the quark-diquark structure of shadow baryons.
However, they do not give  essential contributions to baryon
density, by reason of small $\theta$-charges.

Since H-sector is cooler than the ordinary one, then we have
$n'_B\gtrsim n_B$ by estimate of Ref.~\ct{42}, and:
\be \rho'_B=n'_BM'_B > \rho_B=n_BM_B.  \lb{15r} \ee
Now we can explain the relation (\ref{8AC}), especially if we take
into account the shadow helium mass fraction  (see Ref.~\ct{42}).

Finally, we predict that the energy density of hidden sector is:
\be   \rho' =  \rho_{DE} + \rho_{DM} = \rho_{DE} + \rho'_{B} +
\rho_{CDM}, \lb{16r} \ee
where $\rho_{DE}$ is given by (\ref{10AC}), $\rho'_{B}=n'_BM'_B
\approx 0.17 \rho_c$ and $\rho_{CDM}\approx 0.04 \rho_c$
presumably contains shadow helium.

The energy density of the O-world is;
   \be \rho_M = \rho_B + \rho_{nuclear},   \lb{17r} \ee
where $\rho_B=n_BM_B\approx 0.04 \rho_c$ and the contribution of
ordinary helium and other atoms is much smaller. Then it is
possible to explain the observable result (see Eq.~(\ref{8AC})):
\be  \frac{\Omega_{DM}}{\Omega_{M}} \simeq \frac
{\rho_{DM}}{\rho_M}\simeq
 \frac {\rho'_{B} + \rho_{CDM}}{\rho_B + \rho_{nuclear}}\simeq \frac{0.17 +
0.04}{0.04}\simeq 5.  \lb{17r} \ee

\section{Baryogenesis}

In Ref.~\ct{2} we have presented baryogenesis mechanism in our
cosmological model with superstring-inspired $E_6$ unification. In
this model the $B-L$ asymmetry is produced by the conversion of
ordinary leptons into particles of the hidden sector.

After the non-symmetric reheating with $T_R
> T'_R$, the exchange processes between O- and H-worlds are too
slow, by reason of the very weak interaction between the two
sectors. As a result, it is impossible to establish equilibrium
between them, so that both worlds evolve adiabatically and the
temperature asymmetry ($T'/T < 1$) is approximately constant in
all epochs from the end of the inflation until the present epoch.

The equilibrium between two sectors of massless particles with the
same temperature is not broken by the cosmological expansion, and
the baryon asymmetry (and any charge asymmetry) cannot be
generated in the Universe. However, if there are two components in
the plasma with different temperatures, then the equilibrium is
explicitly broken as long as the temperatures are not equal. In
our case of observed and hidden sectors, the equilibrium never
happens by reason of their essentially different temperatures. In
this case, baryon asymmetry may be generated even by scattering of
massless particles.

In the Bento-Berezhiani model of baryogenesis \ct{38} the heavy
Majorana neutrinos play the role of messengers between ordinary
and mirror worlds. Their model considers the group of symmetry
$G_{SM}\times G_{SM'}$, i.e. the Standard model and its mirror
counterpart. Heavy Majorana neutrinos $N$ are singlets of $G_{SM}$
and $G_{SM'}$ and this is an explanation, why they can be
messengers between ordinary and mirror worlds.

In our model with $E_6$ unification, the $N$-neutrinos belong to
the 27-plet of $E_6$ and $E'_6$, and they are not singlet
particles. But after the breaking
\be E_6\to
 SO(10)\times U(1)_Z\to
 SU(3)_C\times SU(2)_L \times SU(2)_R\times U(1)_X\times
 U(1)_Z  \lb{2C} \ee
in the O-world, and
\be E'_6\to \to SU(6)'\times SU(2)'_{\theta} \to SU(3)'_C\times
SU(2)'_L\times SU(2)'_{\theta}\times U(1)'_X \times U(1)'_Z
\lb{3C} \ee
in the H-world, heavy Majorana neutrinos $N_a$ become singlets of
the subgroups $SU(3)_C\times SU(2)_L \times U(1)_X\times U(1)_Z$
and $SU(3)'_C\times SU(2)'_L\times U(1)'_X \times U(1)'_Z$,
according to Eq.~(\ref{9a}). Therefore, in our model \ct{1}, after
the breaking of $SO(10)$ and $SU(6)'$ and below seesaw scale ($\mu
< M_R=M'_R\sim 10^{10-15}$ GeV), when we have the symmetry groups
$G_{SM}$  and $G_{SM'}\times SU(2)'_{\theta}$, the heavy Majorana
neutrinos $N_a$ again can play the role of messengers between O-
and H-worlds.

Baryon $B$ and lepton $L$ numbers are not perfect quantum numbers.
They are directly related to the seesaw mechanism for light
neutrino masses. $B - L$ is generated in the decays of heavy
Majorana neutrinos, $N$, into leptons $l$ (or anti-leptons $\bar
l$) and the Higgs bosons $\phi$ (which are the standard Higgs
doublets):
\be
      N \to l\phi,\,\,\bar l\bar \phi.  \lb{4C} \ee
In this context, the three necessary Sakharov conditions \ct{65}
are realized in the following way:

1) $B-L$ and $L$ are violated by the heavy neutrino Majorana
masses.

2) The out-of-equilibrium condition is satisfied due to the
delayed decay(s) of the Majorana neutrinos, when the decay rate
$\Gamma(N)$ is smaller than the Hubble rate $H$: $\Gamma(N)< H$,
i.e. the life-time is larger than the age of the Universe at the
time when $N_a$ becomes non-relativistic.

3) CP-violation (C is trivially violated due to the chiral nature
of the fermion weak eigenstates) originates as a result of the
complex $lN\phi$ Yukawa couplings producing asymmetric decay
rates:
\be    \Gamma(N\to l\phi)\neq  \Gamma(N\to \bar l \bar{\phi}),
\lb{5C} \ee
so that leptons and anti-leptons are produced in different amounts
and the $B-L$ asymmetry is generated.

\section{Conclusions}

In this paper we have developed the hypothesis of parallel
existence of the ordinary (O) and hidden (H) sectors of the
Universe. We have constructed a new cosmological model with the
superstring-inspired $E_6$ unification in the 4-dimensional space.
We have assumed that this unification was broken at the early
stage of the Universe into $SO(10)\times U(1)_Z$ -- in the
O-world, and $SU(6)'\times SU(2)'_{\theta}$ -- in the H-world. We
have investigated the breaking mechanism of the $E_6$ unification.
In the O-world this breaking is realized with the Higgs field
$H_{27}$ belonging to the 27-plet, while in the hidden sector the
breakdown of the $E'_6$ unification has come true due to the Higgs
field $H_{351}$ belonging to the 351-plet of the $E'_6$. The
corresponding VEVs are $v=\langle H_{27}\rangle$ and $V=\langle
H_{351}\rangle$. From the beginning, we have assumed that $E'_6$
is the mirror counterpart of the $E_6$. Then the discrete symmetry
$Z_2$ (connected with the mirror parity MP) leads to the
phenomenologically unacceptable wall. Using the simplest model of
inflation with the superpotential $W=\lambda
\varphi(\Phi^2-\mu^2)$, where the field $\varphi$ is the inflaton
and $\Phi$ is the Higgs field, $\lambda$ is a coupling constant
and $\mu$ is a dimensional parameter of the order of the GUT scale
$\sim 10^{18}$ GeV, we avoid this unacceptable wall dominance
assuming the following fine-tuning: $V=V'$, what gives
$\lambda^2\mu^4={\lambda'}^2{\mu'}^4$. Here
$V^{(')}={\lambda^{(')}}^2{\mu^{(')}}^4$ is the energy density of
the tree level potential.

According to our assumptions, there exists the following chains of
symmetry groups:\\ $ E_6\to SO(10)\times U(1)_Z \to SU(4)_C\times
SU(2)_L \times SU(2)_R\times U(1)_Z \to SU(3)_C\times SU(2)_L
\times SU(2)_R\times U(1)_X\times
 U(1)_Z  \to [SU(3)_C\times SU(2)_L\times U(1)_Y]_{{SUSY}}\to
 SU(3)_C\times SU(2)_L\times U(1)_Y $\\
- in the O-world, and\\
$ E'_6 \to SU(6)'\times SU(2)'_{\theta} \to SU(4)'_C\times
SU(2)'_L\times SU(2)'_{\theta}\times U(1)'_Z \to SU(3)'_C\times
SU(2)'_L\times SU(2)'_{\theta}\times U(1)'_X \times U(1)'_Z \to
[SU(3)'_C\times SU(2)'_L\times SU(2)'_{\theta}\times
U(1)'_Y]_{{SUSY}}\to SU(3)'_C\times SU(2)'_L\times
SU(2)'_{\theta}\times U(1)'_Y $\\ - in the H-world.

In contrast to the results of Refs.~\ct{32,33,36,38,39,41,42},
based on the concept of the parallel existence in Nature of the
mirror (M-) and ordinary (O-) worlds described by a minimal
symmetry $G_{SM}\times G'_{SM}$, we assume the existence of
low-energy symmetry group $ G' = SU(3)'_C\times SU(2)'_L\times
SU(2)'_{\theta}\times U(1)'_Y $ in the H-world and the SM symmetry
group in the O-world. This is a natural consequence of different
schemes of the $E_6$-breaking in the O- and H-worlds considered in
Subsection II.C. In comparison with $G_{SM}$, the group $G'$ has
an additional non-Abelian $SU(2)'_{\theta}$ group whose gauge
fields are massless vector particles 'thetons'. These 'thetons'
have a macroscopic confinement radius $1/\Lambda'_{\theta}$. The
estimate given by Refs.\ct{1} confirms the scale
$\Lambda'_{\theta} \sim 10^{-3}$ eV. Assuming the cancellation
between the 'bare' cosmological constant, $\lambda$, and the
vacuum energy stress, $8\pi G\rho_{vac}$, described only by the SM
contributions of the O- and H-worlds (see Sections VI-VIII), we
explain the small value of $\rho_{DE}$, i.e. the observable tiny
CC, only as a result of the $\theta$-fields condensation:
$\rho_{DE} = \rho^{(eff)}_{vac} = {(\Lambda'_\theta)}^4\simeq
(2.3\times 10^{-3}\,\,{\mbox{eV}})^4$.

Taking into account the modern inflationary models with spectral
index $n_s\simeq 1$, we have considered the inflationary
potentials in zero temperature limit and also at the finite
temperature $T$. With this aim, we have used the Coleman-Weinberg
potential (\ref{5i}) for the singlet scalar field $\varphi$. We
have considered in both O- and H-worlds the first order phase
transition when the inflaton starts its slow-rolling towards the
true minimum of the inflationary potential at $\sigma^{(')}=0$,
and gets this minimum at the end of inflation.

We have discussed how the SM-particles were generated at the end
of inflation: the inflaton decays, and the subsequent
thermalization of these decay products generates the SM-particles.
The inflaton $\sigma$ produces gauge bosons: photons, gluons,
$W^{\pm}$, $Z$, and matter fields: quarks, leptons and the Higgs
bosons, while the inflaton  $\sigma'$ produces hidden particles:
shadow photons, gluons and 'thetons', $W'$, $Z'$, theta-quarks
$q_{\theta}$, theta-leptons $l_{\theta}$, shadow quarks $q'$ and
shadow leptons $l'$, scalar bosons $\phi_{\theta}$ and shadow
Higgs fields $\phi'$.

The O- and H-sectors have different cosmological evolutions: they
never had to be in equilibrium with each other. The Big Bang
Nucleosynthesis (BBN) constraints require that H-sector must have
smaller temperature than O-sector: $T'<T$ \ct{42}. The difference
between the O- and H-worlds is described in terms of two
macroscopic parameters: $ x\equiv {T'}/{T}, \quad \beta\equiv
{\Omega'_B}/{\Omega_B}$, where $T(T')$ is O-(H-) photon
temperature of the Universe at present, and $\Omega_B(\Omega'_B)$
is O-(H-)baryons fraction.

We have considered the reheating and radiation in Sec.10 and Big
Bang Nucleosynthesis in Sec.11. During reheating the exponential
expansion, developed by inflation, ceases and the potential energy
of the inflaton field decays into a hot relativistic plasma of
particles. The relativistic fraction is represented by photons and
neutrinos. The radiation is dominant at the early epochs of the
Universe, but it is negligible at present epoch:
$\Omega_r^{(0)}\ll 1$.

The contribution of the H-degrees of freedom to the observable
Hubble expansion rate, which are equivalent to an effective number
of extra neutrinos $\Delta N_{\nu}=6.14\cdot x^4$, is small
enough. In our model: $\omega_r=\Omega_r h^2=4.2\cdot
10^{-5}(1+x^4)\quad (h=H/H_0)$, where the contribution of
H-species is negligible due to the BBN constraint: $x^4\ll 1$.

At the end of inflation the O- and H-sectors are reheated in a
non-symmetric way: $T_R>T'_R$. After reheating, at $T<T_R$, the
exchange processes between O- and H-worlds are too slow (by reason
of very weak interaction between two sectors), and it is difficult
to establish equilibrium between them. As a result, the
temperature asymmetry ($T'/T < 1$) is approximately constant from
the end of inflation until the present epoch.

We have seen that the cosmological evolutions of the early O- and
H-worlds are very different, in particular, when we consider such
crucial epochs as baryogenesis and nucleosynthesis. The BBN epoch
proceeds differently in the O- and H-worlds and predicts different
abundances of primordial elements. For example, due to the
condition $T'<T$ the density of H-photons number is much smaller
than for O-photons: ${n'_{\gamma}}/{n_{\gamma}}=x^3\ll 1$.

The structure formation in the Universe is connected with the
plasma recombination and matter-radiation decoupling (MRD) epochs.
Also the matter-radiation equality (MRE) is important, which is
given by the relation $ 1+z_{eq} = \Omega_m/\Omega_r \simeq
2.4\cdot 10^4\cdot\Omega_m h^2/(1+x^4)$. During the MRD epoch the
most of electrons and protons recombine into neutral hydrogen and
the free electron density essentially diminishes. The MRD
temperature is $T_{dec}\simeq 0.26$ eV what corresponds to the
redshift $1+z_{dec} = T_{dec}/T_{today}\simeq 1100$. In the
H-world we have the MRD temperature $T'_{dec}\simeq T_{dec}$ and
$1+z'_{dec}\simeq x^{-1}(1+z_{dec})\simeq {1100}/{x}$, what means
that in the H-world MRD occurs earlier than in the O-world.

During the recombination epoch the photon scattering rate drops
below the Hubble expansion rate. The H-photon decoupling epoch
coincides with the MRE epoch. At the end of recombination, the
atoms in the Universe are neutral, photons travel freely and the
Universe becomes transparent. The observation of CMB gives a
picture of the Universe at the end of this epoch.

In Sec.12 we have estimated $\rho_M$ and $\rho_{DM}$ in the
framework of our cosmological model. We assume that shadow baryons
and shadow helium, invisible for ordinary photons, give the main
contribution to dark matter (DM). We explain the observable
result: $\Omega_{DM}/\Omega_{M} \simeq  \rho_{DM}/\rho_M \simeq
5$.

Sec.13 is devoted to the baryogenesis mechanism presented in
Ref.~\ct{2}. In our cosmological model with superstring-inspired
$E_6$ unification, the $B-L$ asymmetry is produced by the
conversion of ordinary leptons into particles of the hidden
sector. After the non-symmetric reheating with $T_R
> T'_R$, it is impossible to establish equilibrium
between the O- and H- sectors, and baryon asymmetry may be
generated even by scattering of massless particles. In our model
with $E_6$ unification existing at the early stage of the
Universe, after the breaking of $E_6{(E'_6)}$, heavy Majorana
neutrinos $N_a$ become singlets of the subgroups $SU(3)_C\times
SU(2)_L \times U(1)_X\times U(1)_Z$ and $SU(3)'_C\times
SU(2)'_L\times U(1)'_X \times U(1)'_Z$, and can play the role of
messengers between O- and H-worlds. $B - L$ quantum number is
generated in the decays of heavy Majorana neutrinos, $N$, into
leptons $l$ (or anti-leptons $\bar l$) and the Higgs bosons
$\phi$: $N \to l\phi,\,\,\bar l\bar \phi$. The three necessary
Sakharov conditions, given by Ref.~\ct{65}, are realized in our
model of baryogenesis.

\section*{Acknowledgements} We are grateful to Masud Chaichian
for useful discussions. The support of the Academy of Finland
under the projects no. 121720 and 127626 is acknowledged. L.V.L.
thanks RFBR grant 09-02-08215-3. C. R. Das gratefully acknowledges
a scholarship from Funda\c{c}\~{a}o para a Ci\^{e}ncia e
Tecnologia ref. SFRH/BPD/41091/2007.


\begin{thebibliography}{99}

\bibitem{1}
C.R.~Das, L.V.~Laperashvili, A.~Tureanu, Eur.Phys.J.C {\bf 66}
(2010) 307; arXiv:0902.4874; AIP Conf.Proc. {\bf 1241} (2010) 639;
arXiv:0910.1669.
\bibitem{2}
C.R.~Das, L.V.~Laperashvili, H.B.~Nielsen, A.~Tureanu,
arXiv:1010.2744 [hep-ph], to appear in Phys. Lett. B.
\bibitem{3}
P. Q.~Hung, Nucl. Phys. B {\bf 747} (2006) 55; J. Phys. A {\bf 40}
(2007) 6871; P.Q.~Hung, P.~Mosconi, hep-ph/0611001; M.~Adibzadeh
and P. Q.~Hung, Nucl. Phys. B {\bf 804} (2008) 223; H.~Goldberg,
Phys. Lett. B {\bf 492} (2000) 153; C. R.~Das and
L.V.~Laperashvili, Int. J. Mod. Phys. A {\bf 23} (2008) 1863;
arXiv:0712.1326 [hep-ph;  Phys. Atom. Nucl. {\bf 72} (2009) 377;
arXiv:0712.0253 [hep-ph].
\bibitem{8}
Particle Data Group, C. Amster et al., Phys. Lett. B 667 (2008) 1.
\bibitem{9}
A. G.~Riess et.~al., Astron. J. {\bf 116} (1998) 1009,
astro-ph/9805201; S. J.~Perlmutter et.~al., Nature {\bf 391}
(1998) 51, astro-ph/9712212; Astrophys. J. {\bf 517} (1999) 565,
astro-ph/9812133; C. L.~Bennett et.~al., Astrophys. J. Suppl. {\bf
148} (2003) 1, astro-ph/0302207; D. N.~Spergel et.~al., Astrophys.
J. Suppl. {\bf 148} (2003) 175, astro-ph/0302209; Astrophys. J.
Suppl. {\bf 170} (2007) 377, astro-ph/0603449; P.~Astier et.~al.,
Astron. Astrophys. {\bf 447} (2006) 31, astro-ph/0510447.
\bibitem{14}
A. Riess et al., Astrophys. J. Suppl. 183 (2009) 109;
arXiv:0905.0697; W.L. Freedman et al., Astrophys. J. 704 (2009)
1036; arXiv:0907.4524; R. Kessler at al., arXiv:0908.4274.
\bibitem{17}
P. J. E.~Peebles and A.~Vilenkin, Phys. Rev. D {\bf 59} (1999)
063505, astro-ph/9810509; C.~Wetterich, Nucl. Phys. B {\bf 302}
(1988) 668; S. M.~Carroll, Phys. Rev. Lett. {\bf 81} (1998) 3067,
astro-ph/9806099; Y.~Fujii and T.~Nishioka, Phys. Rev.  D {\bf42}
(1990) 361; B.~Ratra and P. J. E.~Peebles, Phys. Rev. D {\bf 37}
(1988) 3406; Astrophys. J. {\bf 325} (1988) L17.
\bibitem{18}
A. D.~Linde, {\it The New Inflationary Universe Scenario.} In:
``The Very Early Universe", ed. G.W.~Gibbons, S. W.~Hawking and
S.~Siklos, CUP (1983); A. R.~Liddle and D. H.~Lyth, {\it
Cosmological Inflation and Large-Scale Structure} (Cambridge
University Press, Cambridge, 2000); E. J.~Copeland, M.~Sami and
S.~Tsujikawa, Int. J. Mod. Phys. D {\bf 15} (2006) 1753;
hep-th/0603057.
\bibitem{21}
J. H.~Schwarz, Phys. Rept. {\bf 89} (1982) 223; M. B.~Green, Surv.
High. En. Phys. {\bf 3} (1984) 127; M. B.~Green and J. H.~Schwarz,
Phys. Lett. B {\bf 149} 117 (1984); ibid., B {\bf 151} (1985) 21.
\bibitem{23}
D. J.~Gross, J. A.~Harvey, E.~Martinec and R.~Rohm, Phys. Rev.
Lett. {\bf 54} (1985), 502; Nucl. Phys. B {\bf 256} (1985) 253;
ibid., B {\bf 267} (1986) 75; P.~Candelas, G. T.~Horowitz,
A.~Strominger and E.~Witten, Nucl. Phys. B {\bf 258} (1985) 46.
\bibitem{26} M. B.~Green, J. H.~Schwarz and E.~Witten, {\it Superstring theory}
(Cambridge University Press, Cambridge, 1988).
\bibitem{27}
T. D.~Lee and C. N.~Yang, Phys. Rev. {\bf 104} (1956) 254.
\bibitem{28}
I. Yu.~Kobzarev, L. B.~Okun and I. Ya.~Pomeranchuk, Yad. Fiz. {\bf
3} (1966) 1154 [Sov. J. Nucl. Phys. {\bf 3} (1966) 837].
\bibitem{29}
K.~Nishijima and M. H.~Saffouri, Phys. Rev. Lett. {\bf 14} (1965)
205; L. B.~Okun and I. Ya.~Pomeranchuk, Pis'ma Zh. Eksp. Teor.
Fiz. {\bf 1} (1965) 28; [JETP Lett. {\bf 1} (1965) 167]; Phys.
Lett. {\bf 16} (1965) 338; E. W.~Kolb, D.~Seckel, M. S. Turner,
Nature {\bf 314} (1985) 415; Fermilab-Pub-85/16-A, Jan.1985.
\bibitem{32}
Z.~Berezhiani, A.~Dolgov and R. N.~Mohapatra, Phys. Lett. B {\bf
375} (1996) 26; Z.~Berezhiani and R. N.~Mohapatra, Phys. Rev. D
{\bf 52} (1995) 6607; Z.~Berezhiani, Acta Phys. Polon. B {\bf 27}
(1996) 1503; Int. J. Mod. Phys. A {\bf 19} (2004) 3775.
\bibitem{33}
R.~Foot, H.~Lew, and R.R.~Volkas, Phys. Lett. B {\bf 272} (1991)
67; Mod. Phys. Lett. A {\bf 7} (1992) 2567; R.~Foot, Mod. Phys.
Lett. A {\bf 9}(1994) 169; R.~Foot and R.R.~Volkas, Phys. Rev.D
{\bf 55} (1995) 5147; review by R.~Foot, Int. J. Mod. Phys. D {\bf
13}(2004) 2161.
\bibitem{36}
Z.~Berezhiani, D.~Comelli and N.~Tetradis, Phys. Lett. B {\bf 431}
(1998) 286.
\bibitem{38}
L. Bento and Z. Berezhiani, Phys. Rev. Lett. {\bf 87} (2001)
231304;  Fortsch. Phys. {\bf 50} (2002) 489.
\bibitem{39}
Z. Berezhiani, P. Ciarcelluti, D. Comelli and F. L. Villante, Int.
J. Mod. Phys. D {\bf 14} (2005) 107.
\bibitem{41}
Z.~Berezhiani, {\it Through the looking-glass: Alice's adventures
in mirror world}, in: Ian Kogan Memorial Collection ``From Fields
to Strings: Circumnavigating Theoretical Physics'', Eds.
M.~Shifman et al., World Scientific, Singapore, Vol.~3, pp.
2147-2195, 2005; AIP Conf. Proc. {\bf 878} (2006) 195; Eur. Phys.
J. ST {\bf 163} (2008) 271.
\bibitem{42}
Z.~Berezhiani, L.~ Kaufmann, P.~Panci, N.~Rossi, A.~Rubbia,
A.~Sakharov, {\it Strongly interacting mirror dark matter},
CERN-PH-TH-2008-108, May 2008.
\bibitem{50}
B.~Stech, Fortsch. Phys. {\bf 58} (2010) 692; arXiv:1003.0581.
\bibitem{46}
P.~Athron, S.F.~King, D. J.~Miller, S.~Moretti and R.~Nevzorov,
Phys. Rev. D {\bf 80} (2009) 035009; arXiv:0904.2169;
arXiv:0901.1192.
\bibitem{55}
R.~Slansky, Phys. Rept. {\bf 79} (1981) 1.
\bibitem{56}
Taichiro Kugo, Joe Sato, Prog. Theor. Phys. {\bf 91} (1994) 1217;
hep-ph/9402357.
\bibitem{58}
L. B. Okun, Phys. Usp. {\bf 50} (2007) 380; hep-ph/0606202; S. I.
Blinnikov, {\it Notes on Hidden Mirror World}, arXiv:0904.3609
[astro-ph.CO].
\bibitem{60}
L. B.~Okun, JETP Lett. {\bf 31} (1980) 144; Pisma Zh. Eksp. Teor.
Fiz. {\bf 31} (1979) 156; Nucl. Phys. B {\bf 173} (1980) 1.
\bibitem{dvali}
G.~Dvali, Q.~Shafi, R.~Schaefer, Phys.Rev.Lett. {\bf 73} (1994)
1886.
\bibitem{1CC}
A.~Einstein, Sitz. Preuss. Akad. Wiss., (1917) pp. 142- 152,
translated in H.A. Lorentz et al, {\it The Principle of
Relativity}, Dover, New York, (1952) p. 177 .
\bibitem{2CC}
Y. B.~Zeldovich, JETP Lett. {\bf 6} 316 (1967).
\bibitem{3CC}
S.~Weinberg, Rev. Mod. Phys. {\bf 61} 1 (1989).
\bibitem{4CC}
M. T.~Veltman, Phys. Rev. Lett. {\bf 34} (1975) 777.
\bibitem{1MPP}
C. D.~ Froggatt, L. V.~Laperashvili, R. B.~Nevzorov and H.
B.~Nielsen, Phys. Atom. Nucl. {\bf 67} (2004), 582 [Yad. Fiz. {\bf
67} (2004), 601]; arXiv:hep-ph/0310127. Proceedings of 7th
Workshop on 'What Comes Beyond the Standard Model', Bled,
Slovenia, 19-30 Jul 2004; published in *Bled 2004, What comes
beyond the standard models*, pp. 17-27, DMFA-Zaloznistvo,
Ljubljana, 2004; hep-ph/0412208, hep-ph/0411273.
\bibitem{2MPP}
C.~Froggatt, R.~Nevzorov and H. B. Nielsen, Nucl. Phys. B {\bf
743} (2006) 133, hep-ph/0511259; J. Phys. Conf. Ser. {\bf 110}
(2008) 072012; arXiv:0708.2907 [hep-ph].
\bibitem{4MPP}
D. L.~Bennett and H. B.~Nielsen, Int. J. Mod. Phys. A {\bf 9}
(1994), 5155; ibid., A {\bf 14} (1999) 3313; C. D. Froggatt and H.
B. Nielsen, {\it Origin of Symmetries} (World Scientific,
Singapore, 1991); C. D. Froggatt and H. B. Nielsen, Phys. Lett. B
{\bf 368} (1996) 96.
\bibitem{7MPP} L. V. Laperashvili, Yad. Fiz. {\bf 57} (1994) 501
[Phys. Atom. Nucl. {\bf 57} (1994) 471]; C. R.~Das and L.
V.~Laperashvili, Int. J. Mod. Phys. A {\bf 20} (2005) 5911;
arXiv:hep-ph/0503138.
\bibitem{5CC}
D.J.~Shaw, J.D.~Barrow, {\it A Testable Solution of the CC and
Coincedence Problems}, arXiv:1010.4262[gr-qc].
\bibitem{8CC}
E.~Baum, Phys. Lett. B {\bf 133} (1983) 185; S.~Hawking, Phys.
Lett. B {\bf 134} (1984) 403.
\bibitem{14CC}
R. D.~Peccei and H. R.~Quinn, Phys. Rev. Lett. {\bf 38} (1977)
1440.
\bibitem{61}
A.D.~Dolgov, AIP Conf.Proc. {\bf 1116} (2009) 155;
arXiv:0901.2100.
\bibitem{62}
S.R.~Coleman and E.~Weinberg, Phys. Rev. D {\bf 7} (1973) 1888.
\bibitem{63}
P.Q.~Hung, E.~Masso and G.~Zsembinski, JCAP, {\bf 12}(2006) 004;
arXiv: astro-ph/0604063.
\bibitem{65}
A.D.~Sakharov, Pisma Zh. Eksp. Teor. Fiz. {\bf 5} (1967) 32.


\end{thebibliography}
\end{document}